\documentstyle[12pt]{article}

\def \apny#1#2#3{Ann. Phys. (N.Y.) {\bf#1}, #2 (#3)}

\def \cn{Collaboration}
\def \cp89{{\it CP Violation,} edited by C. Jarlskog (World Scientific,
Singapore, 1989)}

\def \hb87{{\it Proceeding of the 1987 International Symposium on Lepton and
Photon Interactions at High Energies,} Hamburg, 1987, ed. by W. Bartel
and R. R\"uckl (Nucl. Phys. B, Proc. Suppl., vol. 3) (North-Holland,
Amsterdam, 1988)}

\def \ibj#1#2#3{~{\bf#1}, #2 (#3)}
\def \ichep72{{\it Proceedings of the XVI International Conference on High
Energy Physics}, Chicago and Batavia, Illinois, Sept. 6 -- 13, 1972,
edited by J. D. Jackson, A. Roberts, and R. Donaldson (Fermilab, Batavia,
IL, 1972)}

\def \ite{{\it et al.}}
\def \jpg#1#2#3{J. Phys. G {\bf#1}, #2 (#3)}
\def \lkl87{{\it Selected Topics in Electroweak Interactions} (Proceedings of
the Second Lake Louise Institute on New Frontiers in Particle Physics, 15 --
21 February, 1987), edited by J. M. Cameron \ite~(World Scientific, Singapore,
1987)}
\def \ky85{{\it Proceedings of the International Symposium on Lepton and
Photon Interactions at High Energy,} Kyoto, Aug.~19-24, 1985, edited by M.
Konuma and K. Takahashi (Kyoto Univ., Kyoto, 1985)}

\def \np#1#2#3{Nucl. Phys. {\bf#1}, #2 (#3)}

\def \plb#1#2#3{Phys. Lett. B {\bf#1}, #2 (#3)}

\def \prd#1#2#3{Phys. Rev. D {\bf#1}, #2 (#3)}
\def \prl#1#2#3{Phys. Rev. Lett. {\bf#1}, #2 (#3)}

\def \ptp#1#2#3{Prog. Theor. Phys. {\bf#1}, #2 (#3)}

\def \si90{25th International Conference on High Energy Physics, Singapore,
Aug. 2-8, 1990}
\def \slc87{{\it Proceedings of the Salt Lake City Meeting} (Division of
Particles and Fields, American Physical Society, Salt Lake City, Utah, 1987),
ed. by C. DeTar and J. S. Ball (World Scientific, Singapore, 1987)}
\def \slac89{{\it Proceedings of the XIVth International Symposium on
Lepton and Photon Interactions,} Stanford, California, 1989, edited by M.
Riordan (World Scientific, Singapore, 1990)}
\def \smass82{{\it Proceedings of the 1982 DPF Summer Study on Elementary
Particle Physics and Future Facilities}, Snowmass, Colorado, edited by R.
Donaldson, R. Gustafson, and F. Paige (World Scientific, Singapore, 1982)}
\def \smass90{{\it Research Directions for the Decade} (Proceedings of the
1990 Summer Study on High Energy Physics, June 25--July 13, Snowmass,
Colorado),
edited by E. L. Berger (World Scientific, Singapore, 1992)}
\def \tasi90{{\it Testing the Standard Model} (Proceedings of the 1990
Theoretical Advanced Study Institute in Elementary Particle Physics, Boulder,
Colorado, 3--27 June, 1990), edited by M. Cveti\v{c} and P. Langacker
(World Scientific, Singapore, 1991)}

\def \zpc#1#2#3{Zeit. Phys. C {\bf#1}, #2 (#3)}
\begin{document}
\leftline{To be submitted to Physical Review D}
\bigskip
\leftline{November 1992}
\vspace{-42pt}
\rightline{Technion-PH-92-40}
\rightline{PITHA 92/39}
\rightline{EFI 92-53}
\medskip
\bigskip
\centerline{\bf METHOD FOR FLAVOR TAGGING IN NEUTRAL B MESON DECAYS}
\bigskip
\centerline{\it Michael Gronau}
\centerline{Department of Physics, Technion}
\centerline{32000 Haifa, Israel}
\medskip
\centerline{\it Alex Nippe}
\centerline{I. Physikalisches Institut, RWTH}
\centerline{Aachen, Federal Republic of Germany}
\medskip
\centerline{\it Jonathan L. Rosner}
\centerline{Enrico Fermi Institute and Department of Physics}
\centerline{University of Chicago, Chicago, IL 60637}
\bigskip

\begin{quote}
A method is proposed for tagging the flavor of neutral $B$ mesons in the study
of CP-violating decay asymmetries. The method makes use of a possible
difference in interactions in $B \pi$ or $B^* \pi$ systems with isospins 1/2
and 3/2, and would be particularly clean if the $I = 1/2$ systems can be
detected as ``$B^{**}$'' resonances.
\end{quote}
\bigskip
\centerline{\bf I.  INTRODUCTION}
\medskip

So far, CP violation has been seen only in decays of neutral kaons.  The
leading contender for description of this effect, a non-trivial phase in the
Cabibbo-Kobayashi-Maskawa (CKM) matrix [1], makes specific predictions for
CP-violating asymmetries in decays of mesons containing a $b$ quark.  These
asymmetries are particularly easily interpreted in decays of neutral $B$ mesons
to CP eigenstates such as $J/\psi K_S$ [2]. In this manner, uncertainties
associated with unknown final state interactions can be circumvented. However,
the identification of the flavor of the initial $B$ has proved to be
non-trivial.  Methods proposed up to now rely on the production of a $B$ or
$\bar B$ in association with a particle of the opposite beauty quantum number,
whose decay serves to tag the flavor of the neutral $B$ of interest.

In the present article, we describe a method of tagging the flavor of a neutral
$B$ meson which makes use of the (strong) interaction of the decaying $b$ or
$\bar b$ quark with other quarks before the time of decay.  The method is
expected to be successful if there is a clearly identifiable difference between
low-mass pion-$B$ interactions in channels with isospin 1/2 and those with
isospin 3/2.  Such a difference would exist, for instance, if there is a
well-defined region of positive-parity pion-$B$ resonances in the mass range
below 5.8 GeV.

The presence of a $\pi^+$ in a low-mass combination with a neutral $B$ meson is
then circumstantial evidence that the $B$ meson is a $B^0$, and not a $\bar
B^0$.  We show here how to convert such circumstantial evidence to a
quantitative measurement.

A method similar to the one we propose, making use of the decay chain $D^{*\pm}
\to \stackrel{\small (-)}{D^0} \pi^\pm$, has been in use for some time for
tagging the flavor of neutral $D$ mesons [3].  In the case of $B$ mesons, there
is not enough energy for the $B^*$ (the $^3S_1$ state of a $b$ quark and a
light antiquark) to decay to the $B$ (the $^1S_0$ state) and a pion.  The $B^*$
always decays to a $B$ and a photon of about 46 MeV. However, it is highly
likely that there exists a region of positive-parity $J = 0,~1$ and 2
resonances corresponding to the P-wave levels of a $b$ quark and a light ($\bar
u$ or $\bar d$) antiquark which decay to $B \pi$ and/or $B^* \pi$.  As a
result, one can identify (on a statistical basis) the flavor of a neutral $B$
meson by making use of its correlation with an appropriately chosen charged
pion.

Since production of positive-parity charmed meson (``$D^{**}$'') resonances
appears to account for about 20 or 30 \% of $D$ meson production in the $e^+
e^-$ continuum [3,4], the corresponding production of ``$B^{**}$'' resonances
at higher $e^+ e^-$ energies, and perhaps at hadron colliders as well, is
likely to be non-negligible.

A general description of the tagging method is given in Sec.~II. One must
measure decays of a neutral $B^0$ or $\bar B^0$ to states of definite flavor in
order to calibrate production rates. The fact that neutral $B^0$ and $\bar B^0$
mesons mix with one another introduces some unavoidable dilution of the
statistical power of this method.

A simplified approach to tagging, described in Sec.~III, is applicable to
charge-symmetric production processes such as $\bar p p$ and $e^+ e^-$
reactions. In this case, the measured asymmetry is related to the desired
quantity by a dilution factor common to all measured asymmetries, which cancels
if one takes their {\it ratio} for two different final states. Such ratios have
been shown to provide useful information about the fundamental CKM parameters
[5].

A special circumstance allows one to calibrate neutral $B$ production using
decays of charged $B$ mesons. When the initial state has zero isospin,
processes involving neutral and charged $B$ mesons can be related to one
another by means of a simple isospin reflection.  This case is described in
Sec. IV.

The method yields useful results only when there are non-trivial correlations
between the charged pion and the decaying $B$.  The most promising example of
such correlations occurs when the charged pion and the $B$ are decay products
of P-wave $b \bar u$ and $b \bar d$ resonances (or their charge conjugates).
We describe the expected behavior of such resonances in Section V, and mention
kinematic circumstances which require one to know the properties of these
resonances rather precisely.  A more general picture under which such
correlations are expected, based on quark fragmentation, is also described.

We conclude in Section VI.
\bigskip

\centerline{\bf II.  GENERAL TAGGING METHOD}
\medskip

The method we propose relies upon the detection of neutral $B$ mesons with
identified flavor ($B^0$ or $\bar B^0$) in conjunction with a pion of
positive or negative charge nearby in phase space, and the detection of
a CP eigenstate $f$ as a $B^0$ or $\bar B^0$ decay product (we do not
know which, {\it a priori}) in conjunction with a similar pion.  We begin
by discussing the states of identified flavor.

A $B^0$ or $\bar B^0$ may be accompanied by a charged pion nearby in
rapidity (equivalently, in a state of low effective mass with the $B^0$ or
$\bar B^0$).  We propose that a $B^0= \bar b d$ is more likely to be
accompanied by a $\pi^+ = \bar d u$, while a $\bar B^0  = b \bar d$ is more
likely to be accompanied by a $\pi^- = d \bar u$. This may be seen either
from a simple picture of fragmentation, or from the likely existence of
positive-parity resonances in the pion-$B$ or pion-$B^*$ systems.  Such
resonances are expected to have isospin 1/2. We shall discuss them more
extensively in Section V.  There we shall specify more precisely how the
accompanying pion is to be chosen.

We define the relative rates of production of $B^0$ and $\bar B^0$
mesons in low-mass combinations with charged pions to be
\begin{equation}
N(\bar B^0 \pi^-) \equiv P_1~~,~~~
N(\bar B^0 \pi^+) \equiv P_2~~,~~~
N(B^0 \pi^+)      \equiv P_3~~,~~~
N(B^0 \pi^-)      \equiv P_4~~~.
\end{equation}
The first and third channels are ``non-exotic,'' and are the ones in which
we might see some resonant enhancement.  The second and fourth channels are
purely I = 3/2, and no such enhancement is expected.  Hence we anticipate
that $P_1 > P_2$ and $P_3 > P_4$.  In the limit of complete resonance
dominance, we would have $P_2 = P_4 = 0$.

Let us denote a flavored state which we know to have come from a $B^0$ by
$T$ (for ``tag'') and the corresponding state for a $\bar B^0$ by $\bar T$.
Examples of states $T$ include $D^- \pi^+$ and $J/\psi K^{*0}$, where the
$K^{*0}$ is seen to decay to $K^+ \pi^-$.  (We are using the usual
convention in which a $B^0$ meson contains a $\bar b$ quark.)  We can
measure four separate correlations of states $T$ or $\bar T$ with charged
pions.  These measurements serve to normalize the production of $B^0$ and $
\bar B^0$ in combination with the pion of either charge. We assume that the
decay rates of $B^0$ to $T$ and $\bar B^0$ to $\bar T$ are equal.  For the
$D^- \pi^+$ and $J/\psi K^{*0}$ final states, this is true in the standard
picture of CP violation [6], where a single amplitude dominates the decay.
The assumption can, of course, be directly verified.

Because of $B^0 -\bar B^0$ mixing, an initial $B^0$ has relative probabilities
$(1-\chi_d)$ and $\chi_d$ of decaying to the states $T$ and $\bar T$,
respectively.  The parameter $\chi_d$ is related to the mass mixing parameter
$x_d \equiv (\Delta m/\Gamma)_d = 0.71 \pm 0.14$ [7] for neutral $B$ mesons by
\begin{equation}
\chi_d = x_d^2/(2 + 2x_d^2) = 0.17 \pm 0.04~~~.
\end{equation}
The relative numbers of $T \pi$ states of various types are then given by
\begin{equation}
N(\bar T \pi^-) = (1 - \chi_d) P_1 + \chi_d P_4~~~,
\end{equation}
\begin{equation}
N(\bar T \pi^+) = (1 - \chi_d) P_2 + \chi_d P_3~~~,
\end{equation}
\begin{equation}
N(T \pi^+)      = (1 - \chi_d) P_3 + \chi_d P_2~~~,
\end{equation}
\begin{equation}
N(T \pi^-)      = (1 - \chi_d) P_4 + \chi_d P_1~~~,
\end{equation}
where we have omitted an overall branching ratio.  These equations can be
solved pairwise for the $P_i$, since $\chi_d$ is very far from 1/2. For
example, we have
\begin{equation}
P_1 = [1 - 2 \chi_d]^{-1} [(1 - \chi_d) N(\bar T \pi^-)
                               - \chi_d N(     T \pi^-)]~~~,
\end{equation}
\begin{equation}
P_4 = [1 - 2 \chi_d]^{-1} [(1 - \chi_d) N(     T \pi^-)
                               - \chi_d N(\bar T \pi^-)]~~~,
\end{equation}
with similar expressions for $P_2$ and $P_3$ but with $\pi^- \to \pi^+$.  The
error on $\chi_d$ itself is expected to decrease in the future, but some
amplification of the experimental errors in $N(T \pi^\pm)$ and $N(\bar T
\pi^\pm)$ will nonetheless occur in the course of inverting the relations
(3)--(6) to obtain the $P_i$.

Now let a state which was produced as $B^0$ at time $t=0$ decay to a CP
eigenstate $f$ with probability $D_f$, while a state which was $\bar B^0$
at $t= 0$ decays to $f$ with probability $\bar D_f$.  Let the numbers of
final states $f \pi^\pm$ (where the charged pion is that referred to above)
be denoted by $N_{f \pm}$. We are interested in the (time-integrated)
asymmetry [6]
\begin{equation}
A(f) \equiv \frac
{\Gamma(B^0_{t=0} \to f) - \Gamma(\bar B^0_{t=0} \to f)}
{\Gamma(B^0_{t=0} \to f) + \Gamma(\bar B^0_{t=0} \to f)}
= \frac{D_f - \bar D_f}{D_f + \bar D_f}~~~.
\end{equation}

For the final state $f = J/\psi K_S$, it has been shown that the asymmetry
$A(f)$ measures the angle $\beta$ of the unitarity triangle (a fundamental
parameter of the CKM matrix) to a very good approximation [8] (for notation
see, e.g., Ref.~[5]):
\begin{equation}
A(J/\psi K_S) = - \frac{x_d}{1 + x_d^2} \sin 2\beta~~~,
\end{equation}
since a single amplitude contributes to each decay $B^0 \to J/\psi K^0$
and $\bar B^0 \to J/\psi \bar K^0$.  There are some questions [8]
as to whether the same is true for the $\pi^+ \pi^-$ final state, but in
the simplest case one has
\begin{equation}
A(\pi^+ \pi^-) =  - \frac{x_d}{1 + x_d^2} \sin 2\alpha~~~,
\end{equation}
where $\alpha$ is another angle in the unitarity triangle.

Now we measure the numbers of CP eigenstates $f$ in conjunction with
charged pions, always taking the same mass range for $f \pi^\pm$ as we took
for the decays to states of identified flavor. Then the number of $f
\pi^\pm$ states will be
\begin{equation}
N_{f+} \equiv N(f \pi^+) = P_3 D_f + P_2 \bar D_f~~~
\end{equation}
\begin{equation}
N_{f-} \equiv N(f \pi^-) = P_4 D_f + P_1 \bar D_f~~~.
\end{equation}
As long as $P_1 P_3 - P_2 P_4 \neq 0$, we can invert Eqs.~(12) and (13) to
find
\begin{equation}
     D_f = (P_1 N_{f+} - P_2 N_{f-})/(P_1 P_3 - P_2 P_4)~~~,
\end{equation}
\begin{equation}
\bar D_f = (P_3 N_{f-} - P_4 N_{f+})/(P_1 P_3 - P_2 P_4)~~~.
\end{equation}
Since it is likely that $P_1 > P_2$ and $P_3 > P_4$, we expect to be able
to perform this operation.  Its validity depends on the existence of
non-trivial correlations between charged pions and neutral $B$ mesons.
These correlations can be searched for experimentally.

The asymmetry $A(f) = (D_f - \bar D_f)/(D_f + \bar D_f)$ is then
\begin{equation}
A(f) = {(P_1 + P_4)N_{f+} - (P_3 + P_2)N_{f-} \over
        (P_1 - P_4)N_{f+} + (P_3 - P_2)N_{f-}}~~~.
\end{equation}
An explicit form in terms of ``tagging'' final states comes from solving
Eqs.~(3)--(6):
\begin{equation}
\left(1 + x_d^2\right) A(f) =
{[N(\bar T \pi^-) + N(     T \pi^-)] N_{f+}
-[N(     T \pi^+) + N(\bar T \pi^+)] N_{f-} \over
 [N(\bar T \pi^-) - N(     T \pi^-)] N_{f+}
+[N(     T \pi^+) - N(\bar T \pi^+)] N_{f-} }~~~.
\end{equation}
This is our central result.
\bigskip

\centerline{\bf III.  CHARGE-SYMMETRIC PRODUCTION}
\medskip

Let us consider a production process such as $p \bar p$ or $e^+ e^-$
annihilation, in which the cross sections for production of $B^0$ and $\bar
B^0$ states should be equal.  In the $p \bar p$ case this follows from the
charge-conjugation invariance of the strong interactions.  In electron-positron
annihilation the production of a $b$ quark is always accompanied by production
of a $\bar b$ quark, and the subsequent fragmentation into hadrons conserves
charge symmetry.

In the present case we have $P_1 = P_3$ and $P_2 = P_4$, and a simpler result
\begin{equation}
A_{\rm obs}(f) \equiv {N_{f+} - N_{f-} \over N_{f+} + N_{f-}} =
{P_1 - P_2 \over P_1 + P_2}~\cdot A(f)
\end{equation}
follows from Eq.~(16). The first factor corrects for the dilution of the
observed effect as a result of the tagging process.  In order that it be
non-zero, we require only that $P_1 \neq P_2$.  As we have mentioned, $P_1 >
P_2$ is most likely for appropriately chosen pions, and $P_2 = 0$ in the limit
that the interaction in the $I = 1/2$ channel is dominant. In terms of
``tagging'' final states, one may write
\begin{equation}
\left(1 + x_d^2\right) A(f) =
\left[ {N(\bar T \pi^-) + N(\bar T \pi^+) \over
        N(\bar T \pi^-) - N(\bar T \pi^+) } \right]
\cdot A_{\rm obs}(f) ~~~.
\end{equation}
for the case of charge-symmetric production.

Since the required number of events to see an observed asymmetry $A_{\rm obs}$
at the level of $S$ standard deviations is $(S/A_{\rm obs})^2$ [6], this number
is proportional to a factor $(P_1 + P_2)^2/(P_1 - P_2)^2 > 1$. This factor is
to be compared with ones involved in tagging via the associated $B$, which
typically involve a branching ratio to a leptonic or other final state.  There,
it is the inverse branching ratio which governs the required number of events.

Even if the dilution factor is unknown, it cancels out if we study two {\it
different} final states $f$ and $f'$:
\begin{equation}
{N_{f+} - N_{f-} \over N_{f+} + N_{f-}}/
{N_{f'+} - N_{f'-} \over N_{f'+} + N_{f'-}} = A(f)/A(f')~~~.
\end{equation}
The ratio of decay asymmetries for the final states $\pi^+ \pi^-$ and $J/\psi
K_S$, for example, provides interesting information on the parameters in the
Cabibbo-Kobayashi-Maskawa matrix, if the reservations expressed in Ref.~[8] can
be dealt with.  In that case, we have just
\begin{equation}
A(\pi^+ \pi^-)/A(J/\psi K_S) = \sin 2 \alpha / \sin 2 \beta~~~,
\end{equation}
leading to simple geometric constraints on the unitarity triangle [5].
\bigskip

\centerline{\bf IV.  ISOSCALAR PRODUCTION}
\medskip

\leftline{\bf A.  Cases of isoscalar production}
\medskip

We shall be able to make use of isospin reflection symmetry whenever the
initial state leading to a $B \pi$ resonance has isospin zero.  Cases in which
this can occur include the following:

{\it 1.  The reaction $e^- e^+ \to b \bar b$.}  If the current (acting through
a virtual photon or virtual $Z^0$) produces the $b \bar b$ pair directly, that
pair is of course produced with zero isospin.  The subsequent fragmentation to
hadrons also conserves isospin.

{\it 2.  Hadronic collisions involving projectiles with $I = 0$,} such as
deuterium-deuterium, carbon-carbon, oxygen-oxygen, or $^{40}$Ca-$^ {40}$Ca
collisions.

{\it 3.  Production of the $b \bar b$ state and its fragmentation products
(including the pions in the $B \pi$ resonances) from an initial gluonic state},
as occurs in any perturbative QCD description of hadronic $b \bar b$
production.  This condition can be violated if the pions do not come from
fragmentation of the $b$ quarks.
\bigskip

\leftline{\bf B. Isospin Relations}
\medskip

When isospin reflection $I_3 \to -I_3$ is a good symmetry, one has the
following relations between final states involving $B$'s and associated pions:
\begin{equation}
N(B^- \pi^+) = N(\bar B^0 \pi^-) = P_1~~,~~
N(B^- \pi^-) = N(\bar B^0 \pi^+) = P_2~~,~~
\nonumber
\end{equation}
\begin{equation}
N(B^+ \pi^-) = N(     B^0 \pi^+) = P_3~~,~~
N(B^+ \pi^+) = N(     B^0 \pi^-) = P_4~~~,
\end{equation}
where we have used the definitions of Sec.~II.
One must be careful that isospin splittings between charged and neutral
$B^{**}$ resonances are not so large as to cause measurable effects, but this
appears highly unlikely.  The situation is quite different in the case of the
decays $D^* \to D \pi$, where some channels are actually closed as a result of
such splittings.

The result of Sec.~II for decay asymmetries now may be transcribed directly.
Expressed in terms of ratios of measured numbers of events, it is
\begin{equation}
A(f) = {N_{f+}[N(B^- \pi^+) + N(B^+ \pi^+)] - N_{f-}[N(B^+ \pi^-)
+ N(B^- \pi^-)] \over   N_{f+}[N(B^- \pi^+) - N(B^+ \pi^+)] + N_{f-}
[N(B^+ \pi^-) - N(B^- \pi^-)]
}~~~.
\end{equation}
One needs non-trivial $B$-pion correlations which differ in exotic $(I=3/2)$
and non-exotic channels in order for the relation to be useful.

As in Sec.~III, the assumption of a charge-symmetric production process allows
the above relation to be expressed in terms of the product of a dilution
factor, now given in terms of charged $B$ rates, and an observed asymmetry.
Explicitly, we have
\begin{equation}
A(f) = {N(B^- \pi^+) + N(B^+ \pi^+) \over N(B^- \pi^+) - N(B^+ \pi^+)}
{}~\cdot~{N_{f+} - N_{f-} \over N_{f+} + N_{f-}}~~~.
\end{equation}
This is the case in $e^+ e^-$ annihilation.
\bigskip

\centerline{\bf V.  CORRELATIONS OF $B$ MESONS AND PIONS}
\medskip

\leftline{\bf A.  Resonance spins and decay channels}
\medskip

A $b$ quark and a light antiquark can form P-wave positive-parity resonances
with $J = 0,~ 1$, and 2.  The $J=0$ and $J=2$ states are pure spin-triplets,
while two physical $J=1$ states are expected to be linear combinations of
spin-singlet and spin-triplet states with definite values of light-quark total
angular momentum (spin + orbital angular momentum) [9,10].  The allowed decay
channels are:
\begin{equation}
B(J = 0) \to B \pi~~,~~~
B(J = 1) \to B^* \pi~~,~~~
B(J = 2) \to B \pi,~B^* \pi~~~.
\end{equation}

Detection of the soft photon emitted in $B^*$ decay could help identify the
P-wave states with maximum efficiency.

\bigskip

\leftline{\bf B. Mass estimates: extrapolation from charmed mesons}
\medskip

In the limit of heavy-quark symmetry [10], the energy required to excite a
light antiquark bound to a heavy quark should be independent of the mass of the
heavy quark.  Accordingly, we shall use the masses of the observed P-wave
charmed mesons to estimate those of the corresponding excited $B$ mesons.

There are candidates for P-wave charmed mesons at 2420 and 2460 MeV [4]. It is
likely that the state at 2420 MeV corresponds to one or both of the expected
$J=1$ levels, since it decays only to $D^* \pi$.  The state at 2460 MeV
probably corresponds to the $J=2$ level, since it is seen to decay both to $D
\pi$ and to $D^* \pi$.

We will not know the fine-structure splitting in the P-wave charmed meson
multiplet until all four states have been identified. However, it is likely
that the spin-averaged mass $\bar M_P(D)$ of the P-wave charmed mesons lies at
or below that of the $J=1$ candidate at 2420 MeV.  We estimate the mass
difference associated with a P-wave excitation by comparing this value with the
spin-averaged mass of S-wave charmed mesons:
\begin{equation}
\bar M_S(D) \equiv [3M(D^*) + M(D)]/4 = 1973~{\rm MeV}~~~,
\end{equation}
where we have used averages over isospin splittings.  Consequently, we
estimate that
\begin{equation}
\bar M_P(D) - \bar M_S(D) \stackrel{<}{\sim} 450~{\rm MeV}~~~.
\end{equation}
To reduce the uncertainty on this number, it would be very helpful to detect
the $J = 0$ state, decaying only to $D \pi$.

The spin-averaged mass of S-wave $B$ mesons is
\begin{equation}
\bar M_S(B) \equiv [3M(B^*) + M(B)]/4 = 5315~{\rm MeV}~~~.
\end{equation}
If $\bar M_P(B) - \bar M_S(B) \stackrel{<}{\sim} 450$ MeV as suggested by the
corresponding value for charmed mesons, {\it we expect a region of P-wave $B
\pi$ or $B^* \pi$ resonances lying below 5.8 GeV.}

The detection of a photon from the decay $B^* \to B \gamma$ may be difficult,
since the photon will only have 46 MeV in the $B^*$ center of mass.  If a
resonance of mass 5.8 GeV or less decays to a $B^*$ meson and a pion, and the
photon in the decay of the $B^*$ meson is not detected, the resulting $B \pi$
system will also have an effective mass less than 5.8 GeV, but should still be
confined to a rather narrow mass interval since the photon is so soft.

The fine-structure splittings in heavy-quark - light-quark systems are
characterized by two distinct scales [9,10].  First, the light quark and the
orbital angular momentum are coupled up to a total angular momentum $J_{\rm
light} = 1/2$ or 3/2.  The mass difference associated with $J_{\rm light} =
1/2$ or 3/2 will not change when we go from $D$ to $B$ mesons.  It could be
considerable; we don't know, since we have only seen two states [$D(2420)$ and
$D(2460)$] so far. Second, $J_{\rm light}$ couples to the spin of the heavy
quark.  When $J_{\rm light} = 1/2$, we get states of total spin $J = 0,~1$,
while when $J_{\rm light} = 3/2$ we get $J = 1,~2$. The splitting of the two
states with a given $J_{\rm light}$ should behave as the inverse of the heavy
quark mass.  If the $D(2420)$ and $D(2460)$ are both states of $J_{\rm light} =
3/2$, for example, their $B$ meson analogues may be closer together in mass.
This proximity could be an advantage in reducing backgrounds.
\bigskip

\leftline{\bf C.  Isospin considerations}
\medskip

Let us henceforth ignore the soft photon which may be emitted in $B^*$ decay,
and speak of resonances in the $B \pi$ system as standing for both $B \pi$ and
$B^* \pi$.  These resonances should occur only in $I=1/2$ (``non-exotic'') and
not in $I = 3/2$ (``exotic'') channels.  Similar behavior is noted for
resonances involving strange particles.  Non-exotic mesonic channels correspond
to states which can be formed of a quark and an antiquark, while exotic mesons
require at least two quarks and two antiquarks.  No exotic mesons have been
observed up to now.

We expect resonances in the channels $B^- \pi^+$, $\bar B^0 \pi^-$, $B^+
\pi^-$, and $B^0 \pi^+$, but not in the channels $B^- \pi^-$, $\bar B^0 \pi^+$,
$B^+ \pi^+$, or $B^0 \pi^-$. All channels with a neutral pion should contain
resonances, but with strength half of that in the isospin-related channels
involving charged pions.

It is most likely that accidental exotic combinations of a pion and a $B$ can
be avoided when the fragments of the corresponding antiparticle are far away in
rapidity.  Thus, the method we propose may not be particularly useful for
production of a $b \bar b$ pair near threshold.  The reaction $e^- e^+ \to Z^0
\to b \bar b$ would seem to be ideal for present purposes, if sufficient
statistics can be obtained. The program we propose includes
measurements which will {\it check} whether non-exotic and exotic $B \pi$
combinations show a different mass spectrum.
\newpage

\leftline{\bf D.  Fragmentation of $b$ and $\bar b$ quarks}
\medskip

Another argument in favor of non-trivial correlations between pions and $B$
mesons, at least when the $B-\pi$ system is of low effective mass, may be
presented in the language of quark fragmentation.  At the same time, this
argument exposes a source of potential dilution of the correlation.

Let us consider a $b$ quark to fragment to a $\bar B^0 = b \bar d$ meson.
Somewhere not far from the $B^0$ in phase space there should then exist a $d$
quark, which is the partner of the $\bar d d$ pair which has been produced in
the fragmentation process.  This $d$ quark is more likely to give rise to a
$\pi^-$ than to a $\pi^+$.  However, its probability for generating a $\pi^+$
in a low-mass combination with the $\bar B^0$ is non-zero.  For example, the
$d$ quark could fragment to a $\rho^0$ meson, which then decays to $\pi^+
\pi^-$.

For such reasons, the detection of correlations between pions and $B$ mesons
may require experimental study rather than mere theoretical speculation.  There
may be particularly favorable regions of phase space which we have not
anticipated for which the difference between non-exotic and exotic $B \pi$
channels is most pronounced.

The present method appears to be a special case of a more general approach.
The basic idea behind this generalization is the fact that the charge of the
leading quark can propagate through and become visible in the jet containing
the $B$ meson. Such a jet-charge method has been employed in the identification
of light quarks, for instance in Ref.~[11].  First, this can result in a
definite charge relation between the whole jet and the $B$ flavor. Second, one
might expect differences in the shape of properly choosen kinematical
variables, now taking all pion fragments of the jet into account. Such
variables are, for example, the mass of the  $B \pi$ system, or the momentum or
transverse momentum of the pion in a suitable frame.  In a manner equivalent to
the calibration process described above, one can extract these distributions
directly from the data and use them as input for a (multidimensional) fit or
even a neural network analysis.  We have described a method which places the
most emphasis on the leading pion since those differences are expected to be
most obvious for it.
\bigskip

\leftline{\bf E.  Some kinematic considerations}
\medskip

Some simple examples show that it may not be trivial to reduce combinatorial
backgrounds in establishing $B \pi$ correlations.  These examples underscore
the importance of a detailed understanding of resonances in the $B \pi$ and
$B^* \pi$ systems.

Let us consider the effective mass of a $B \pi$ system in two reference frames:
 one in which the pion is at rest, and one in which the pion has 300 MeV of
energy and is traveling transverse to the $B$.

In the first frame, the $\pi B$ effective mass does not exceed the value of 5.8
GeV (our proposed upper limit for the lowest positive-parity resonances) until
the $B$ energy exceeds about 21 GeV.  A $B$ produced in $e^+ e^-$ annihilations
at the $Z^0$ mass has an average energy of 30 -- 35 GeV, but a hadronically
produced $B$ is unlikely to have such an energy with regard to any of its
fragments since the hadronic production processes favors $b \bar b$ final
states not far above threshold.  The pions formed as a result of the filling of
the rapidity gap between $b$ and $\bar b$ may be fairly soft with respect to
{\it both} the $b$ and the $\bar b$.

The second frame is probably a more realistic representation of a centrally
produced pion in the $b \bar b$ center of mass in either a high-energy $e^+
e^-$ or hadronic collision.  Here, the $B$ energy corresponding to an effective
$B \pi$ mass of 5.8 GeV is only about 9 GeV.  Accordingly, accidental low-mass
combinations of a $B$ and a ``wrong'' pion are expected to be less frequent.

In order to solve the system of equations in Sec.~II for $D_f$ and $\bar D_f$,
one must be able to see a difference between exotic and non-exotic channels.
It may be necessary to make rather strict cuts on $B \pi$ systems such that
they have a high probability of having originated in the lowest positive-parity
resonances. Identification of the soft photon in $B^*$ decay could help in
making use of the specific masses of the $J=1$ resonances, and could also
enhance the signal from the decay of the $J=2$ state, but may not be essential,
since its omission would shift and broaden the $B \pi$ mass distribution only
slightly.
\bigskip

\centerline{\bf VI.  CONCLUSIONS}
\medskip

We have suggested a way to identify the flavor of a $B^0$ or $\bar B^0$
decaying to a CP eigenstate by means of a pion forming a low-mass combination
with the neutral $B$ meson.  The most natural source of this combination is a
band of positive-parity $J=0,~1$, and 2 resonances lying somewhere below 5.8
GeV.  If this band is rather narrow, the presence of a $\pi^+$ in this
combination is circumstantial evidence in favor of the neutral $B$ having been
a $B^0$, while a $\pi^-$ suggests an initial $\bar B^0$.

We have not discussed a corresponding method for tagging $B_s$ decays. Isospin
forbids the decay $B_s^{**} \to B_s \pi$, while $B^{**} \to B_s K$ is likely to
be kinematically forbidden.  A method of tagging a $B_s$ with an
accompanying kaon in a jet has been suggested in Ref.~[12].

We have proposed several means of converting ``circumstantial evidence'' to
quantitative measurements which can be interpreted in terms of CP-violating
asymmetries.  The number of events required to observe a given asymmetry is
proportional in certain simplified cases to a factor $(P_1 + P_2)^2/(P_1 -
P_2)^2$, where $P_1$ and $P_2$ are the relative probabilities for non-exotic
and exotic low-mass $B \pi$ correlations.  What is needed at present is an
experimental study of the nature of these correlations in various production
configurations, to see if they are strong enough to provide the needed
information.  The potential for tagging a neutral $B$ via a particle
nearby it in phase space, rather than via the decay (e.g., to leptons)
of an associated $b$-flavored hadron, has interesting implications for lepton
identification and detector size in future experiments searching for CP
violation in the $B$ meson system.

\newpage

\centerline{\bf ACKNOWLEDGMENTS}
\medskip
We thank Sheldon Stone for helpful comments. J. L. R. is grateful to the CERN
Theory Group and the Physics Department of the Technion for extending gracious
hospitality. This work was supported in part by the United States-Israel
Binational Science Foundation under Research Grant Agreement 90-00483/2, by the
Fund for Promotion of Research at the Technion, by the German Bundesministerium
f\"ur Forschung und Technologie, and by the U. S. Department of Energy under
Grant No. DE FG02 90ER 40560.
\bigskip


\centerline{\bf REFERENCES}
\medskip

\begin{enumerate}

\item[{[1]}] M. Kobayashi and T. Maskawa, \ptp{49}{652}{1973}.

\item[{[2]}] A. B. Carter and A. I. Sanda, \prl{45}{952}{1980}; \prd{23}{1567}
{1981}; I. I. Bigi and A. I. Sanda, \np{B193}{85}{1981}.

\item[{[3]}] S. Stone, HEPSY-1-92, April, 1992, to be published in {\it
Heavy Flavors,} edited by A. J. Buras and M. Lindner, World Scientific,
Singapore, 1992.

\item[{[4]}] H. Albrecht \ite~(ARGUS \cn), \plb{221}{422}{1989};
\ibj{231}{208}{1989}; \ibj{232}{398}{1989}; J. C. Anjos \ite~(Fermilab E691
\cn), \prl{62}{1717}{1989}; P. Avery \ite~(CLEO \cn), \prd{41}{774}{1990}.

\item[{[5]}] P. F. Harrison and J. L. Rosner, \jpg{18}{1673}{1992}.

\item[{[6]}] I. Dunietz and J. L. Rosner, \prd{34}{1404}{1986}; I. Dunietz,
\apny{184}{350}{1988}.

\item[{[7]}] P. Drell, Rapporteur's talk at XXVI International Conference on
High Energy Physics, Dallas, TX, August, 1992.

\item[{[8]}] M. Gronau, \prl{63}{1451}{1989}; SLAC-PUB-5911, September,
1992, to appear in Phys.~Lett.~B.

\item[{[9]}] A. De R\'ujula, H. Georgi, and S. L. Glashow,
\prl{37}{398}{1976}; \ibj{37}{785}{1976}.

\item[{[10]}] N. Isgur and M. B. Wise, \prl{66}{1130}{1991}; M. Lu, M. B. Wise,
and N. Isgur, \prd{45}{1553}{1992}.

\item[{[11]}] D. Decamp \ite~(ALEPH \cn), \plb{259}{377}{1991}; \plb{284}{177}
{1992}.

\item[{[12]}] A. Ali and F. Barreiro, \zpc{30}{635}{1986}.
\end{enumerate}
\end{document}